\documentclass[review]{elsarticle}

\usepackage{lineno,hyperref}
\modulolinenumbers[5]
\usepackage{graphicx}
\usepackage{amsmath}
\usepackage{epstopdf}
\usepackage{braket}
\usepackage{float}
\journal{Physics Letters A}

\begin{document}

\begin{frontmatter}

\title{Topological phases in Bi/Sb planar and buckled honeycomb monolayers}

\author[mymainaddress,mysecondaryaddress]{N. Nouri\corref{mycorrespondingauthor}}
\cortext[mycorrespondingauthor]{Corresponding author}
\ead{nafise.nour@sci.ui.ac.ir}

\author[mymainaddress]{M. Bieniek}
\author[mymainaddress]{M. Brzezi\'{n}ska}
\author[Mohsenadresss,Mohsenadresss2]{M. Modarresi}
\author[Borujeniadresss]{S. Zia Borujeni}
\author[mysecondaryaddress]{Gh. Rashedi}
\author[mymainaddress]{A. W\'{o}js}
\author[mymainaddress]{P. Potasz}

\address[mymainaddress]{Department of Theoretical Physics, Faculty of Fundamental Problems of Technology, Wroc\l{}aw University of Science and Technology, 50-370 Wroc\l{}aw, Poland}
\address[mysecondaryaddress]{Department of Physics, Faculty of Sciences, University of Isfahan, Isfahan 81746-73441, Iran}
\address[Mohsenadresss]{Laboratory of Organic Electronics, Department of Science and Technology, Linkoping University, 60174 Norrkoping, Sweden}
\address[Mohsenadresss2]{Department of Physics, Ferdowsi University of Mashhad, Mashhad, Iran}
\address[Borujeniadresss]{Department of Mathematics, Tehran Central Branch, Islamic Azad University, Tehran, Iran}

\begin{abstract}
We investigate topological phases in two-dimensional Bi/Sb honeycomb crystals considering planar, buckled, freestanding and deposited on a substrate structures. We use the multi-orbital tight-binding model and compare results with density functional theory methods. We distinguish topological phases by calculating topological invariants, analyze edge states of systems in a ribbon geometry and by looking at their entanglement spectra. We show that weak coupling to the substrate of buckled and planar crystals is sufficient to lead to a transition to the $Z_2$ topological insulator phase. Topological crystalline insulator (TCI) phase exhibits a pair of edge states in the semi-infinite geometry and in the entanglement spectrum. Transport calculations for TCI phases show robust quantized conductance even in a presence of symmetry-breaking disorder.
\end{abstract}


\end{frontmatter}


\section{Introduction}
Novel topological phases have gained an immense interest of both experimentalists and theorists in the current decade\cite{TI,TI1,TI2,TI3,TI4,TCI1,TCI2,TCI3,TCI4,TCI5}. They are attractive subjects due to potential technological applications in spintronic and quantum computing devices\cite{application}. Topological insulators exhibit the energy gap inside the bulk and conduction channels at the edges, inherently protected against certain types of scattering. Quantum spin Hall (QSH) systems are two-dimensional (2D) representatives of the family of $Z_2$ topological insulators 
protected by time-reversal symmetry\cite{QSH}. The topologically nontrivial energy gap is opened by spin-orbit coupling (SOC), which is a characteristic of heavy elements. QSH systems were experimentally observed in thickness-tunable quantum wells and honeycomb-like systems based on groups of IV \cite{QSH1,QSH2,QSH3}, II-VI \cite{QSH4,QSH6}, III-V \cite{QSH7,QSH8,QSH9,QSH10,QSH11} and V \cite{dir111ofBi1,QSH12,QSH13,QSH14} elements including heavy atoms like bismuth or antimony. Recently, thin films of topological insulators protected by crystalline symmetries were recognized and dubbed topological crystalline insulators \cite{TCI1,TCI2,TCI3,TCI4,TCI5}.

Bi and Sb crystals were extensively studied in the context of their topological properties. Murakami has predicted that Bi bilayer is $Z_2$ topological insulators with characteristic helical edge modes propagating in opposite directions \cite{dir111ofBi1}. This was later confirmed experimentally by scanning tunneling microscopy measurements \cite{expBi1,expBi2,expBi3,science}. Several authors investigated robustness of topological properties of Bi(111) and Sb(111) bilayers and few bilayer crystals\cite{edge1Bi,edge2Bi, BiSb1, BiSb2, BiSb3, BiSb4}. Sb(111) thin films with less than four bilayers were shown to be topologically trivial\cite{QSH13}. Transitions between topologically trivial and nontrivial phases can be induced by structure modifications involving chemical methods \cite{Freitas}, artificial variation of spin-orbit coupling in Bi \cite{edge1Bi,SOC}, strain \cite{QSH14,strainsb,dir111ofSb1,HuangPRB88} or interaction with a substrate\cite{helicaledge1,substrate1BiSb,substrate2BiSb,science,HuangPRB88}. Recently, TCI phase in flat Bi and Sb honeycomb layers was predicted \cite{TCI3}. In the presence of strain, buckled structures become completely flat, which leads to forming bismuthene (for Bi crystals) and antimonene (for Sb). 

In this work, we investigate different topological phases in planar and buckled Bi and Sb two-dimensional honeycomb layers using multi-orbital tight-binding (TB) and compare results with density functional theory (DFT) calculations. We analyze whether TB method can be used as a complementary tool to characterize these crystals, as one of its advantage is studies of larger systems. We distinguish different topological phases for freestanding structures and deposited on a substrate. These phases are identified by computing topological invariants, looking at band structures in a ribbon geometry and analyzing entanglement spectra (ES). We focus mainly on characteristic features of TCI phase and study its topological protection against scattering by calculating conductance in a presence of crystal-symmetry breaking disorder. 

The paper is organized as follows. In Section \ref{sec:Meth} we introduce the methodology. We compare tight-binding method with density functional theory calculations during buckled-flat transitions of freestanding systems in Section \ref{sec:vacuum}. In Section \ref{sec:ZNR}, we analyze the effect of the interaction with the substrate. In Section \ref{sec:Entang} we characterize topological phases using entanglement spectrum and in Section \ref{sec:transport} transport properties of TCI phase are studied. We conclude the results in Section \ref{sec:conclusion}.
\section{\label{sec:Meth}Methods}

\begin{figure}
\hspace{0cm}
\centerline{\includegraphics[width=1.0\columnwidth]{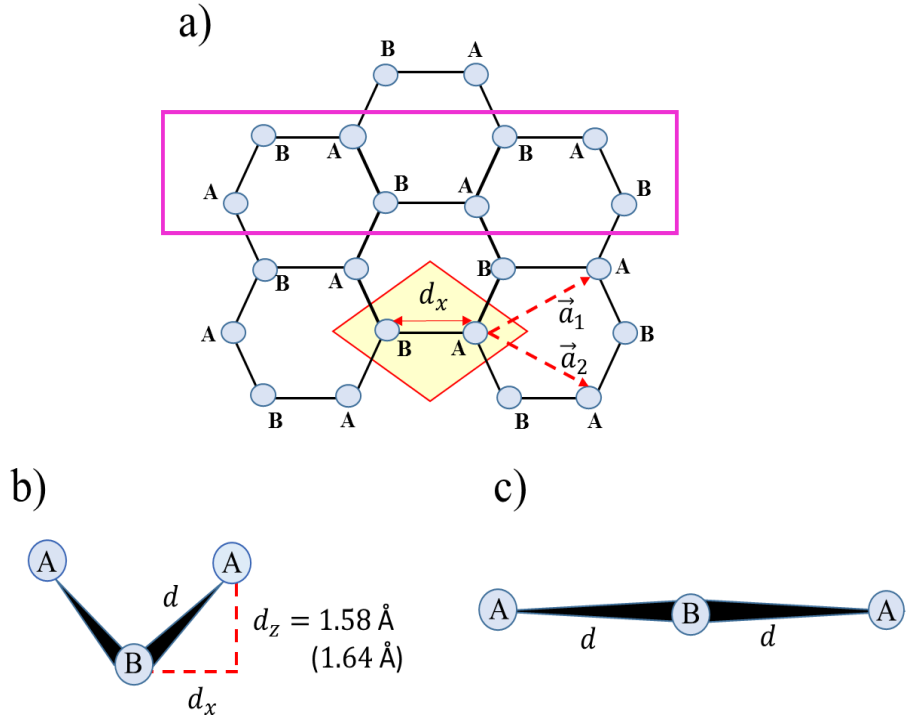}}
\caption{A honeycomb crystal. (a) Top view. $\vec{a}_1$ and $\vec{a}_2$ are the lattice vectors. Orange rhomboid denotes a unit cell of infinite layer, whereas pink rectangle corresponds to the zigzag nanoribbon with periodic boundary conditions in a vertical direction. The atoms from a infinite layer unit cell are labeled as $A$ and $B$. Side views of (b) bismuth and antimony bilayers, together with (c) bismuthene and antimonene, respectively. In our TB method, we set the out of plane distortion to ${d_z} = 1.58\,\,${\AA} for bismuth and ${d_z} = 1.64\,\,${\AA} for antimony. $d$ is the bond length between the nearest neighboring atoms.}
\label{fig:fig1}
\end{figure}

2D crystals Bi and Sb 2D are schematically shown in Fig. \ref{fig:fig1}(a). A hexagonal unit cell contains two atoms, $A$ and $B$. Freestanding Bi and Sb honeycomb layers have the lowest energy when two atoms are displaced in a vertical direction to a lattice plane, thus they are usually called bilayers \cite{dir111ofBi1}. While this displacement slightly differs in a literature depending on type of calculation\cite{HuangPRB88,HiraharaPRL109,LiuPRL107}, we take $d_z=1.58$ {\AA} for Bi and $d_z=1.64$ {\AA} for Sb. We will consider transitions between buckled Bi and Sb bilayers, Fig. \ref{fig:fig1}(b), to flat honeycomb crystals, bismuthene and antimonene (see Fig. \ref{fig:fig1}(c)), caused by an external uniform strain.

\subsection{\label{sec:tb}Tight-binding model}
We use four-orbital ($s,p_x,p_y,p_z$) TB method with parametrization introduced by Liu and Allen \cite{LiuAllenPRB} for bulk bismuth and antimony. The inter-atomic hopping up to the next nearest-neighbors and the atomic spin-orbit coupling (SOC) are parametrized with the Slater-Koster approach\cite{SlaterKoster}. Therefore, we can write Hamiltonian as
\begin{equation}
\begin{array}{l}
H = \sum\limits_{\alpha ,\sigma ,R} {\left[ {\left| {\alpha ,\sigma ,R} \right\rangle {E_\alpha }\left\langle {\alpha ,\sigma ,R} \right|} \right]} \\ +

\sum\limits_{\alpha ,\beta ,\sigma ,R,R'} {\left[ {\left| {\alpha ,\sigma ,R} \right\rangle V_{\alpha \beta }^{\rm I}\left\langle {\beta ,\sigma ,R'} \right| + H.c} \right]} \\ +

\sum\limits_{\alpha ,\beta ,\sigma ,R,R''} {\left[ {\left| {\alpha ,\sigma ,R} \right\rangle V_{\alpha \beta }^{{\rm I}{\rm I}}\left\langle {\beta ,\sigma ,R''} \right| + H.c} \right]} \\ +
\frac{\lambda}{3} \sum\limits_{\alpha ,\beta ,\sigma ,\sigma ',R} {\left[ {\left| {\alpha ,\sigma ,R} \right\rangle \,\vec L\cdot\vec \sigma \left\langle {\beta ,\sigma ',R} \right| + H.c} \right]}, 
\end{array}
\end{equation}\label{1}
where $\left\{ {\alpha ,\,\beta } \right\}$ label orbital $\left\{ {s,\,{p_x},\,{p_y},{p_z}} \right\}$ and spin $\left\{ {\sigma ,\sigma '} \right\}$ degrees of freedom, $R' (R'')$ denote atomic positions of the nearest ${\rm I}$\,(next-nearest {\rm I}{\rm I}) neighbors to atom localized at $R$. ${E_\alpha }$ corresponds to the on-site energies and ${V_{\alpha \beta }}$ are Slater-Koster two-center integrals between ${\alpha}$ and ${\beta }$ orbitals. The last term describes the spin-orbit coupling (SOC) with strength $\lambda$. TB parameters are listed in Table. \ref{tab:t1}. $1/3$ factor is introduced to renormalized atomic SOC strength $\lambda$ in order to obtain correct SOC splitting of the valence band \cite{Chadi}. Buckled-flat transitions are modeled by linearly decreasing $d_z$ and at the same time linearly increasing lattice constants of Bi and Sb bilayers up to the values corresponding to completely flat bismuthene and antimonene, with $a=5.35$\,{\AA} and $a=5.00$\,{\AA}, respectively. According to Ref. \cite{science}, the SiC substrate is effectively described by shifting $p_z$ orbitals away from the low-energy sector.
\begin{table}
\centering
\begin{tabular}{|c|c|c||c|c|c|}
\hline 
Parameter & Bi & Sb & Parameter & Bi & Sb \\ 
(eV) & & & (eV) & & \\ \hline
E$_{s}$ & -10.906 & -10.068 & ${\left( {pp\sigma } \right)_{\rm I}}$ & 1.854 & 2.342 \\ 
E$_{p}$ & -0.486 & -0.926 & ${\left( {pp\pi } \right)_{\rm I}}$ & -0.600 & -0.582 \\
${\left( {ss\sigma } \right)_{\rm I}}$ & -0.608 & -0.694 & ${\left( {pp\sigma } \right)_{\rm II}}$ & 0.156 & 0.352 \\
${\left( {sp\sigma } \right)_{\rm I}}$ & 1.320 & 1.554 & $\lambda$ & 1.5 & 0.6 \\
\hline
a ($\textnormal{\AA}$) & 4.53 & 4.30 & d$_z$ ($\textnormal{\AA}$) & 1.58$^{*}$ & 1.64$^{**}$ \\
d$_x$ ($\textnormal{\AA}$) & 2.62 & 2.48 & & &\\
\hline
\end{tabular} 
\vspace{0.5cm}
\caption{Bismuth and antimony two-center hopping integrals taken from Refs. \cite{dir111ofSb1}$^{**}$, \cite{parameter}$^{*}$, \cite{latticeconstant}. $a$ is a lattice constant, and $d_x$ and $d_z$ denote parallel and perpendicular (buckling) distance between nearest neighbor atoms in a honeycomb lattice, indicated in Fig. \ref{fig:fig1}(b).}
\label{tab:t1}
\end{table}

\subsection{\label{sec:dft}Density functional theory}
We study the atomic configurations and electronic properties of relaxed and strained 2D Sb and Bi crystals in the DFT framework. The DFT calculations are done within the generalized gradient approximation (GGA) and the Perdew-Burke-Ernzerhof (PBE) \cite{perdew1996generalized} exchange correlation function. The core electrons were model using the norm-conserving pseudo potentials. The cut-off energy for the plane wave expansion and charge density calculations are set to 75/85 and 750/850 Ry for Sb/Bi 2D layers. The distance between layers is 15 $\mbox{\AA}$ to avoid interaction between adjacent image layers. We apply the biaxial tensile strain which saves the hexagonal shape of the relaxed unit cell. The first Brillouin zone integration is performed in the Monkhorst-Pack algorithm \cite{monkhorst1976special} using a 15$\times$15$\times$1 k-grid for relaxation of strained atomic configurations. In the relaxation, the total force on each atom in the final configuration is less than 0.001 (Ry/au). We start DFT relaxation from a completely flat structure and a buckled one to compare the final energy and find the lowest energy configuration. For band structure calculations which include the SOC, we used fully-relativistic pseudo potentials and a fine mesh of 25$\times$25$\times$1 for k-grid. All DFT calculations presented in this article were performed using the Quantum-Espresso package \cite{QUANTUMESPRESSO}.

The relaxed Bi and Sb honeycomb lattices are buckled and with lattice constants $a=4.45$ and $a=4.13$ $\mbox{\AA}$, respectively. The equilibrium buckling of relaxed structures are 1.64 $\mbox{\AA}$ for Sb and 1.69 $\mbox{\AA}$ for Bi 2D planes. The resulted atomic configurations are in a good agreement with previous reports \cite{TCI6, akturk2016single}. Obtained values differ from chosen tight-binding parameters by less than $10\%$, and we verify that this discrepancy do not affect results in a qualitative way.

\subsection{\label{sec:ES}Entanglement spectrum}
Quantum entanglement measures have emerged as valuable tools in investigating topologically nontrivial phases of matter. Starting from a system in a ribbon geometry, we divide it into two spatially separated sub-regions denoted by $A$ and $A^c$. The reduced density matrix for the region $A$ captures non-local correlations and can be represented in a form of $\rho_A = e^{ - H_A } /Z_A$, with $Z_A$ being the normalization factor, since $Tr \rho_A = 1$. $H_A$ is called the entanglement Hamiltonian, hence the entanglement spectrum (ES) is defined as a set of eigenvalues of $H_A$ denoted by $\lbrace \xi \rbrace$. If the non-interacting fermionic system is considered, then $H_A$ is just the Hamiltonian restricted to the sites within the one subsystem. 

For free fermion lattice systems, the reduced density matrix can be obtained from the one-particle correlation function \cite{Peschel}
\begin{equation}
C^{\alpha \beta}_{ij}= \textrm{Tr} \left( \rho_A c^{\dagger}_{i\alpha} c_{j\beta} \right) = \braket{GS | c_{i \alpha}^{\dagger} c_{j \beta} | GS},
\label{corr}
\end{equation}
where $\lbrace i, j \rbrace$ are lattice indices within the subsystem $A$, $\lbrace \alpha, \beta \rbrace$ correspond to orbital or spin degrees of freedom and the expectation value $\langle \ldots \rangle$ is taken in the ground state $\ket{GS}$. If translational invariance is present in the system, the full Hamiltonian $H$ can be written in the momentum space and the many-body ground can be expressed in terms of Bloch states $\ket{GS} = \prod_{n k} a^{\dagger}_{n k} \ket{0}$. Here $k$ is the conserved momentum and $n$ runs over the occupied bands. In that case the correlation matrix becomes $k$-dependent, $C \rightarrow C(k)$ and therefore can be evaluated for each point in the Brillouin zone separately.

Correlation matrix $C$ can be regarded as a spectrally flattened physical Hamiltonian $H$ with eigenvalues $\xi$, which are bounded by $0$'s and $1$'s. If system is in a topologically nontrivial phase, a continuous set of intermediate eigenvalues spectrally connecting $0$'s and $1$'s called the spectral flow is exhibited \cite{Hughesinv, Vishinv}. Our aim is to observe whether distinct features between TI and TCI phases are observed. 
\subsection{\label{sec:transport}Transport calculations}
We consider two-terminal geometry with semi-infinite leads attached to the left and the right edge of the scattering region. We use the Landauer approach for the differential conductance $G=e^{2}h^{-1} T$, where $T$ is transmission matrix calculated using recursive Green's functions approach, as explained in our previous work \cite{edge1Bi}. Disorder effects were introduced using Anderson model with on-site values chosen randomly from a uniform distribution $\left[ {-W/2,W/2} \right]$ (with $W$ being the disorder strength) and the conductance averaged over 100 samples with the different impurity realizations. All transport results were calculated for the nanoribbon scattering region size $N=50\times100$ atoms without the inclusion of strain due to its negligible effect on a qualitative picture.

\section {\label{sec:vacuum}Comparison between density functional theory and tight-binding results for freestanding structures}
\begin{figure}
\hspace{0cm}
\centerline{\includegraphics[width=1.0\columnwidth]{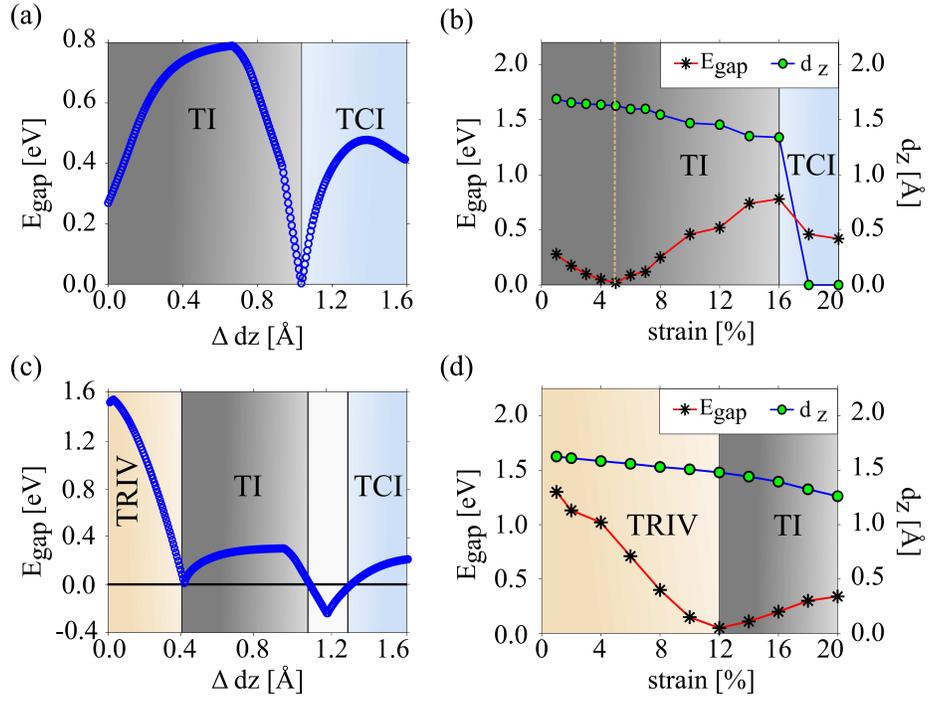}}
\caption{The energy band gap $E_{gap}$ as a function of buckling $\Delta dz = {d_z}_0 - {d_z}$ for structures in the vacuum for (a) and (b) Bi, and (c) and (d) Sb 2D crystals. ${d_z}_0$ refers to the buckling of Bi and Sb bilayers without strain. (a,c) Tight-binding and (b,d) DFT results. Distinct topological phases are marked by different colors.}
\label{fig:fig2}
\end{figure}
Methods within first-principles calculations are commonly used in order to study electronic properties of Bi and Sb honeycomb 2D crystals \cite{dir111ofBi1,QSH14,expBi1,expBi3,science,edge1Bi,edge2Bi,BiSb1, BiSb2, BiSb3, BiSb4,QSH13,Freitas,SOC,strainsb,dir111ofSb1,HuangPRB88,TCI3}. We justify applicability of TB model by comparison results with our DFT calculations. We analyze the effect of biaxial strain leading to a transition between buckled and flat structures. Within TB method, bilayer structures correspond to $\Delta dz=0$, where $\Delta dz = {d_z}_0 - {d_z}$ with ${d_z}_0$ given in Table \ref{tab:t1} and ${d_z}$ as a buckling for strained systems. Fig. \ref{fig:fig2} shows corresponding phase diagrams obtained within TB and DFT for Bi in (a) and (b), and for Sb in (c) and (d). The $Z_2$ topological invariant is calculated from the parities of filled bands at four time-reversal invariant momentum (TRIM) points\cite{TI2}. Qualitatively, phase diagrams look similarly within both methods. Unstrained Bi bilayer ($\Delta dz=0$) is a $Z_2$ TI\cite{dir111ofBi1, parameter, helicaledge,helicaledge1,edge1Bi}. A continuous change of $d_z$ within TB model leads to a transition to TCI phase, which occurs at $\Delta dz =1.04$\,{\AA} and is expected in bismuthene \cite{TCI3}. Within DFT methods, we notice that after a critical value of strain, $16\%$ in this case, the system relaxes to the flat structure exhibiting TCI phase, which has the lowest energy. One can also notice the energy gap closing point around $4\%$ strain, however we verify that the system remains in TI phase. This is not observed in TB results. Agreement between phase diagrams obtained by TB and DFT methods is observed for Sb 2D honeycomb crystal. Strain induces a transition from a trivial phase to TI at around $12\%$ in DFT and at $\Delta dz\sim 0.4$ in TB, and to TCI phase for $\Delta dz\sim 1.3$ in TB, the phase expected for antimonene \cite{TCI3}. For TB results, a semi-metallic phase is observed between TI and TCI phases at around $\Delta dz\sim 1.2$, however this corresponds to strain values inducing relaxation to a flat structure and thus is not seen in DFT calculations. The energy gap evolution within DFT results from Fig. \ref{fig:fig2}(b) and (d) are also comparable with results from Ref. \cite{QSH14}.

\section{\label{sec:ZNR}Substrate effect within tight-binding method}
\begin{figure}
\hspace{0cm}
\centerline{\includegraphics[width=1.0\columnwidth]{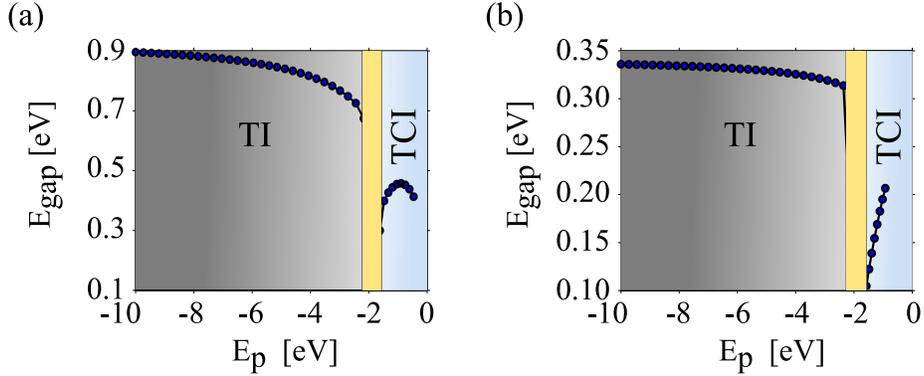}}
\caption{The energy band gap $E_{gap}$ as a function of interaction with a substrate modeled by changing of energy $E_p$ of $p_z$ orbitals for (a) bismuthene and (b) antimonene. $E_p=-0.486$ eV in (a) and $E_p=-0.926$ eV in (b) correspond to freestanding bismuthene (antimonene), see Table \ref{tab:t1}, and $E_p=-10$ eV to structures deposited on the substrate. The yellow areas refer to $E_p$ values with not well defined Fermi energy (see the main text), coinciding with transition regions from TCI to TI phases.}
\label{fig:fig3}
\end{figure}

We investigate the effect of interaction between Bi and Sb 2D crystals with SiC substrate within TB method. Samples deposited on the substrate are coupled to it mainly through orbitals perpendicular to the layer, $p_z$ orbitals in this case. This can be modeled by shifting up the orbitals energy $E_p$ from a low energy sector \cite{science}. We note that this effective model describing the substrate effect unable to determine correctly the Fermi energy at intermediate values of $E_p$, when energy bands continuously shifts in energy with $E_p$. We exclude the region from the phase diagram where the band consisted mainly of $p_z$ orbitals crosses the Fermi energy, the yellow areas in Fig. \ref{fig:fig3}. In Fig. \ref{fig:fig3}(a) and (b), energy gaps $E_{gap}$ as a function of $E_p$ are shown for bismuthene and antimonene, respectively. $E_p=-0.486$ eV ($E_p=-0.926$ eV) corresponds to freestanding bismuthene (antimonene), see Table \ref{tab:t1}, and $E_p=-10$ eV to structures deposited on the substrate. TB model predicts bismuthene (antimonene) $E_{gap}\sim 0.9$ eV ($E_{gap}\sim 0.34$ eV), comparable with DFT results from Ref. \cite{substrate1BiSb} with structures on top of SiC under different tensile strain. Additionally, we show that quite weak coupling to the substrate is sufficient to a transition from TCI to TI phase, which occurs around $E_p\sim-2.5$ eV in both structures. After this phase transition, the energy gap is stable and only slightly affected by coupling strength with the substrate. Energy gap $E_{gap}$ in bismuthene on the substrate ($E_p=-10$ eV) is almost three times larger than the gap in antimonene, compare energy scales in Fig. \ref{fig:fig3}(a) and (b). This can be related to $2.5$ times larger SOC strength in Bi (see Table \ref{tab:t1}). Energy gap of bismuthene on the substrate $E_{gap}=0.9$ eV is also over three times larger than the energy gap of bismuth (111) bilayer, $E_{gap}=0.25$ eV, which was noticed in Ref. \cite{science}.

\begin{figure}
\centerline{\includegraphics[width=1.0\columnwidth]{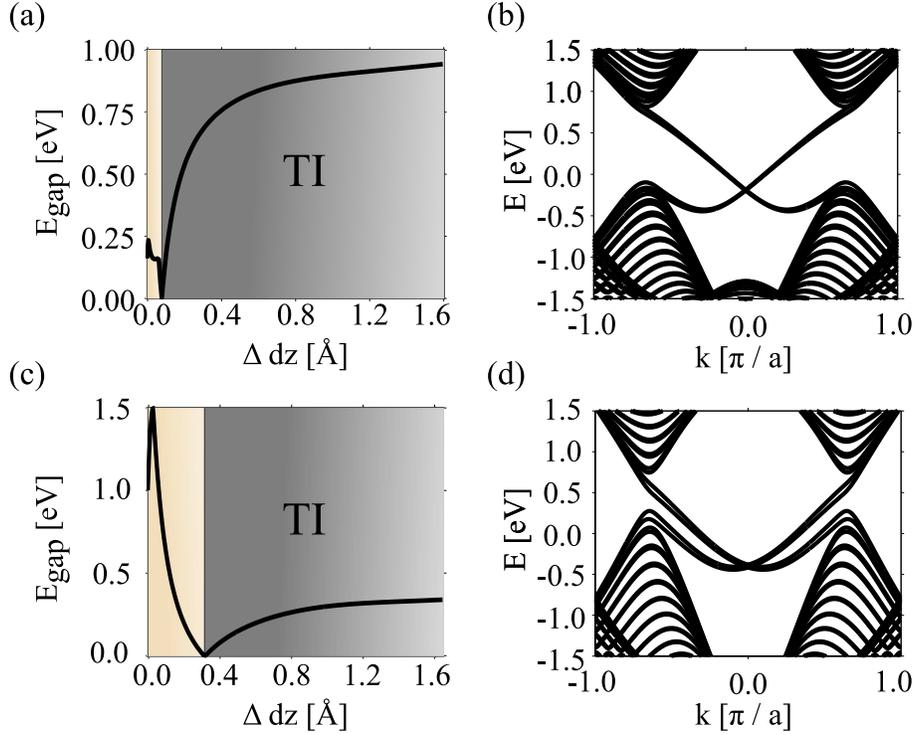}}
\caption{The energy band gap $E_{gap}$ as a function of buckling $\Delta dz = {d_z}_0 - {d_z}$ for structures on the SiC substrate for (a) Bi and (c) Sb 2D crystals. ${d_z}_0$ refers to the buckling of Bi and Sb bilayers without strain. In (a) and (c) topological phases are marked by different colors. (b) and (d) The band structures of zigzag nanoribbon of Bi and Sb crystals within TI phase with buckling of ${d_z}$=$1\,\,${\AA}. Interaction with a substrate splits double degeneracy of each branch of edge states.}
\label{fig:fig4}
\end{figure}

We consider also a transition between bismuth (antimony) bilayer and bismuthene (antimonene) by applying external strain for structures deposited on the substrate. The coupling to the substrate of buckled structure is modeled by shifting the energy $E_{p}$ of $p_z$ for one atom from a unit cell. The second atom has the energy of $p_z$ orbital $E_p=-10$ eV, as it is fully coupled to the substrate for all strain values. In Fig. \ref{fig:fig4}(a) and (c) we show that bilayers ($\Delta dz =0$) deposited on a substrate are within a trivial insulator phase. One can see that a very small strain corresponding to $\Delta dz \sim 0.1$ for bismuth and $\Delta dz \sim 0.1$ for antimony, induces a transition to TI phase with the energy gap monotonically increasing to the largest value for flat systems, bismuthene and antimonene, the right limits of $E_p$ in Fig. \ref{fig:fig4}(a) and (c) respectively. In Fig. \ref{fig:fig4}(b) and (d) we show corresponding band structures of systems in a ribbon geometry for $d_z=1$ {\AA}. A pair of edge states crosses the energy gap, as is expected for TI phase. The degeneracy of these edge states is removed due to inversion-symmetry breaking - two atoms from a unit cell are inequivalent for a buckled structure as they coupled with different strength to the substrate.

\section{\label{sec:Entang}Entanglement spectra of Bi and Sb zigzag nanoribbons}
Topological properties of the system can be confirmed by analyzing the structure of the entanglement spectrum. The presence of spectral flow in ES is associated with nontrivial band topology of the system. The number of intersecting branches of edge states in the band structure of a system in a ribbon geometry allows one to distinguish between TI and TCI phases, compare Fig. \ref{fig:fig4}(b) and (d) (TI) with Fig. \ref{fig:fig6}(a) and (c) (TCI). Similar features can be also observed in entanglement spectra.
\begin{figure}
\centerline{\includegraphics[width=1.0\columnwidth]{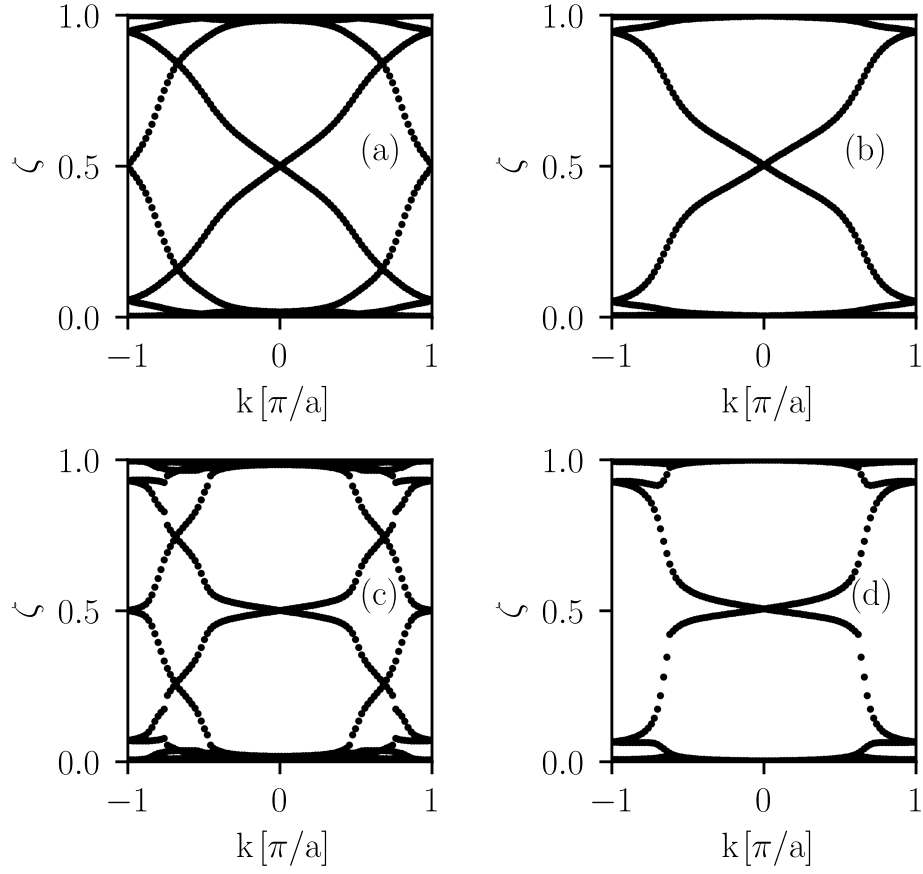}}
\caption{Single-particle entanglement spectra of (a) bismuthene and (c) antimonene in a vacuum, which correspond to TCI phase. (b) Bi and (d) Sb bilayers on the substrate exhibit TI phase. One and two pairs of intersecting branches of edge states crossing the energy gap are characteristic features of TI and TCI phases, respectively.}
\label{fig:fig5}
\end{figure} 

In Fig. \ref{fig:fig5}, we present the single-particle ES for (a) bismuthene and (c) antimonene in a vacuum, together with results for (b) Bi and (d) Sb bilayers on a substrate. Two pairs of intersecting branches are exhibited in TCI phase. It is in analogy to the edge modes in the band structures from Ref. \cite{TCI3} and Fig. \ref{fig:fig6} in Section \ref{sec:transport}. In Figs \ref{fig:fig5} (b) and (d), only one pair of modes spectrally connecting $0$'s and $1$'s is noticed and we identify these spectra with TI phase.

\section{\label{sec:transport}Transport properties}
\begin{figure}
\centerline{\includegraphics[width=1.0\columnwidth]{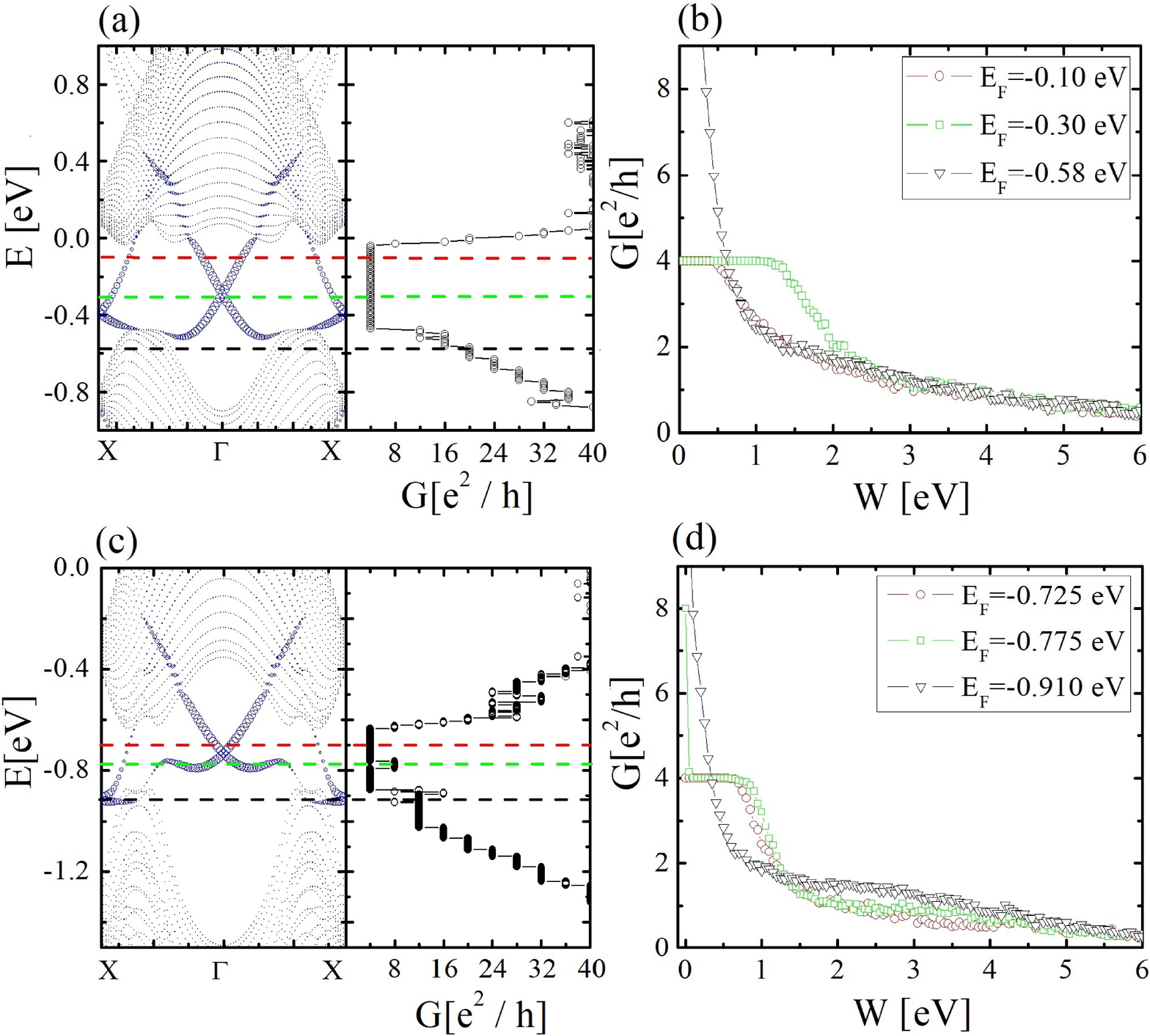}}
\caption{Electronic and transport properties of freestanding bismuthene (a) and (b), and antimonene (c) and (d), nanoribbons in TCI phase. (a) and (c) Left: Ribbon band structures in the 1st 1D Brillouin zone. Size of the blue circles represents level of localization of the wavefunction of a given state on the two atoms on both edges of the nanoribbon. Right: The conductance $G$ in clean system. (b) and (d) The conductance for three different Fermi energies, marked in (a) and (b) by horizontal color lines, as a function of the Anderson disorder strength $W$.}
\label{fig:fig6}
\end{figure}

We focus on the topological protection of transport through edge states in zigzag-type nanoribbons of bismuthene and antimonene in a vacuum, corresponding to systems within TCI phase. In both, bismuthene in Fig. \ref{fig:fig6}(a) and antimonene in Fig. \ref{fig:fig6}(c), we observe four branches of in-gap edge states with edge localization denoted by a size of blue circles. The edge states extend deeply into nanoribbon conduction band, which is novel with respect to previously studied Bi and Sb nanoribbons in TI region \cite{edge1Bi}. Transport in the clean samples with the Fermi energy $E_{F}$ in the energy gap shows expected conductance equal to $4{e^2}/h$ for bismuthene. Antimonene edge states have nonlinear dispersion and eight edge states cross the Fermi level at some energies, e.g. around the Fermi energy $E_{F}\approx -0.8$ eV, \ref{fig:fig6}(c).
Fig. \ref{fig:fig6} (b) and (d) present results of transport calculations in the disorder samples for three different Fermi energies in both bismuthene and antimonene. For $E_F=-0.10$ eV in bismuthene in Fig. \ref{fig:fig6}(b) one can observe protection against backscattering up to the disorder strength $W\approx 0.5$ eV. This critical value of $W$ increases with the lowering of the Fermi energy to $E_F=-0.3$ eV and can be explained by a larger energetic distance to scattering channels from the conduction band states and a higher level of localization at the edges of the system, larger blue circles in Fig \ref{fig:fig6}(a). A critical disorder strength $W$ is smaller in a case of antimonene, which can be related to nonlinear dispersion of edge states and possible scattering within edge channels and also the smaller energy gap in comparison to bismuthene. Analogously to TI nanoribbons, we do not observe any protection against scattering for transport with Fermi energies within both conduction and valence bands, e.g. the latter case represented by $E_F= -0.58$ eV in Fig. \ref{fig:fig6}(a) and $E_F= -0.91$ eV in Fig. \ref{fig:fig6}(c). However, within the energy gap region, while Anderson type of disorder breaks a crystal symmetry, robust conductance is still expected.

\section{\label{sec:conclusion}Summary and discussion}
In summary, we have studied topological phases in Bi and Sb planar and buckled honeycomb crystals using multi-orbital TB method and compared the results with our and previous DFT calculations. We have shown that TB method correctly predicts topological phases in these systems. Weak coupling to the substrate of buckled and planar crystals is sufficient to lead to a transition to $Z_2$ topological insulator phase. We have also proved that entanglement spectra can distinguish between TCI and TI phases which is revealed by two and one pair of branches of entanglement energies in a spectral flow, respectively. We have analyzed also TCI phase in the context of topological protection against scattering during transport through edge channels. Robust quantized conductance is observed even in a presence of a symmetry breaking Anderson disorder.

\section{acknowledgments} 
The authors acknowledge partial financial support from National Science Center (NCN), Poland, grant Maestro No. 2014/14/A/ST3/00654. Our calculations were performed in the Wroc\l{}aw Center for Networking and Supercomputing. 

\bibliography{reference}

\end{document}